\documentclass[aps,prb,showpacs,twocolumn,superscriptaddress]{revtex4-1}
\usepackage{graphicx}

\begin{document}
\title{Dirac Point and Edge States in a Microwave Realization of Tight-Binding Graphene-like Structures}

\author{U.~Kuhl}
\affiliation{Fachbereich Physik, Philipps-Universit\"{a}t Marburg, Renthof 5,
35032 Marburg, Germany}
\affiliation{Laboratoire de Physique de la Mati\`{e}re Condens\'{e}e, CNRS UMR 6622, Universit\'{e} de Nice Sophia-Antipolis, 06108 Nice, France}
\author{S.~Barkhofen}
\affiliation{Fachbereich Physik, Philipps-Universit\"{a}t Marburg, Renthof 5,
35032 Marburg, Germany}
\author{T.~Tudorovskiy}
\affiliation{Fachbereich Physik, Philipps-Universit\"{a}t Marburg, Renthof 5,
35032 Marburg, Germany}
\author{H.-J.~St\"{o}ckmann}
\affiliation{Fachbereich Physik, Philipps-Universit\"{a}t Marburg, Renthof 5,
35032 Marburg, Germany}
\author{T.~Hossain}
\affiliation{Laboratoire de Physique de la Mati\`{e}re Condens\'{e}e, CNRS UMR 6622, Universit\'{e} de Nice Sophia-Antipolis, 06108 Nice, France}
\author{L.~de Forges de Parny}
\affiliation{Laboratoire de Physique de la Mati\`{e}re Condens\'{e}e, CNRS UMR 6622, Universit\'{e} de Nice Sophia-Antipolis, 06108 Nice, France}
\author{F.~Mortessagne}
\affiliation{Laboratoire de Physique de la Mati\`{e}re Condens\'{e}e, CNRS UMR 6622, Universit\'{e} de Nice Sophia-Antipolis, 06108 Nice, France}

\date{\today}

\begin{abstract}
  We present a microwave realization of finite tight-binding graphene-like structures.
  The structures are realized using disks with a high index of refraction.
  The disks are placed on a metallic surface while a second surface is adjusted atop the discs, such that the waves coupling the disks in the air are evanescent, leading to the tight-binding behavior. In reflection measurements the Dirac point and a linear increase close to the Dirac point is observed, if the measurement is performed inside the sample. Resonances due to edge states are found close to the Dirac point if the measurements are performed at the zigzag-edge or at the corner in case of a broken benzene ring.
\end{abstract}

\pacs{42.70.Qs, 73.22.-f, 71.20.-b, 03.65.Nk}
\maketitle
Due to its electronic shell structure carbon forms structures such as diamonds, fullerenes and nanotubes, which all have very specific and fascinating properties, mechanically as electronically. Another realization is graphene, a one-atom-thick allotrope of carbon, which has a structure of the honeycomb lattice, i.\,e.\ two combined triangular lattices. A common feature of all sheet-like carbon structures including graphene is $sp^2$ hybridization leaving one $p_z$ orbital for the bonding. This is the starting point for the tight-binding approximation used in many theoretical descriptions of graphene including the disseminating works by Wallace \cite{wal47} and Semenoff.\cite{sem84} Due to the specific symmetry of the lattice the gapless band structure has a conical singularity at two $k$ vectors,\cite{wal47} called nowadays the Dirac points. The name comes from the fact that the reduced equation around the Dirac points corresponds to the Dirac equation of a massless particle resulting in a band structure of the form of two cones touching at the tips.\cite{nov05,cas09} Recently it was shown that this single layer can be realized and investigated experimentally.\cite{nov04,gei07}
The main ingredient for this relation is the symmetry. Thus a realization via a photonic crystal \cite{sou93,lou05} was proposed.\cite{sep08} In addition in two microwave experiments signatures of the Dirac point had been found in the transmission in an triangular set-up.\cite{zan10,bit10} Both experiments correspond to {\em open} scattering arrangements. Graphene, on the other hand, corresponds to a {\em closed} system because of the tight-binding between nearest neighbors preventing an escape to the outer world. We consider the aspect of symmetry and the aspect of tight-binding with mainly nearest neighbor interaction equally important.\cite{ben09b} This is realized by a microwave experiment using dielectric disks with a high index of refraction in a hexagonal lattice. The disks are weakly coupled by evanescent modes as will be explained later.

So far microwave experiments have been used to show different kinds of effects originally coming from condensed matter physics such as the Hofstadter butterfly,\cite{kuh98b} localization in disordered systems,\cite{cha00,lau07} and transport in case of correlated disorder.\cite{kuh08a}

In the proposed microwave setup it is feasible to investigate defects, edges, disorder etc., as the realized ``graphene'' is a macroscopic object. Thus it is possible to study problems experimentally which are not accessible in real graphene at the moment. Additionally one can have a look into the system and study e.\,g.\ the density of states (DOS) or transport properties as well. In this paper we will concentrate on the local DOS  (LDOS) and its characteristics at edges or corners.

\begin{figure}
\includegraphics[width=.8\columnwidth]{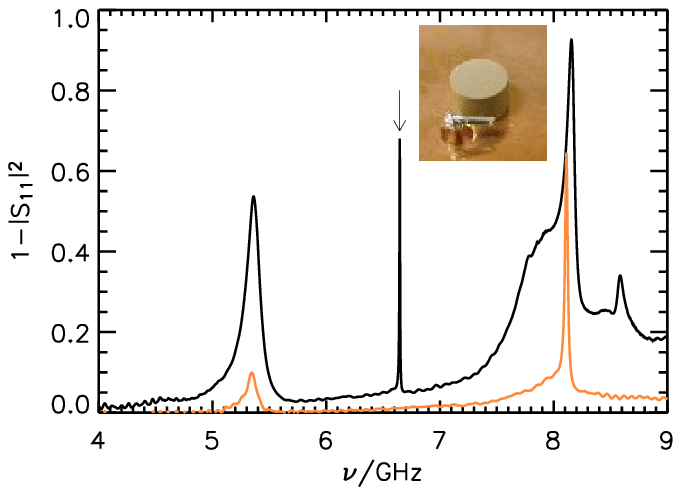}
\vspace{-0.6cm}

\caption{\label{fig:SingleDisc}
(color online) Spectrum of a single disk for the TE antenna (dark, black) and the TM antenna (light, yellow). The TE$_1$ resonance of the disc, which is used for further investigations is marked.
The inset shows the TE antenna with a single disk on the metallic copper plate. The top plate is not shown.
\vspace*{-0.5cm}}
\end{figure}

As a starting point we investigate a single disc, then the coupling of two of them and proceed via a benzene ring to graphene flakes. We measure the reflection $S_{11}$ of an antenna situated close to a disk of $r_d$=4\,mm radius, $h_d$=5\,mm height and an index of refraction $n_d\approx$~6. The antenna is bended (see inset of Fig.~\ref{fig:SingleDisc}) to excite both transverse magnetic (TM) and transverse electric (TE) modes whereas a standard dipole antenna can only excite the TM modes. The disk is placed on a metallic plate (see inset of Fig.~\ref{fig:SingleDisc}), a second plate is covering the whole setup at height $h$ = 16\,mm. The same disks have been used in Ref.~\cite{lau07} to find localization in disordered microwave structures. In contrast to previous experiments the top plate is not touching the disks but it is 11\,mm above the disc. The distance is chosen such that the resonance is still sharp but the eigenfrequency is not sensitive on small deviations of the distance to the top plate. In Fig.~\ref{fig:SingleDisc} the measured spectrum, more precisely $1-|S_{11}|^2$, of a single disk is shown. The dark black curve presents the spectrum measured by the antenna shown in the inset coupling both to the TE and TM modes whereas the light curve was measured with a pure dipole antenna only coupling to the TM mode. The TE$_{01}$-resonance which is used solely for the further investigations is marked. For all measurements we took care that the resonances below and above the investigated frequency range were still separated from the frequency range of interest. The eigenfrequency of the TE resonance decreased with increasing height of the plate, demonstrating that it is a TE$_{01}$-resonance of the disc.

\begin{figure}
\includegraphics[width=.9\columnwidth]{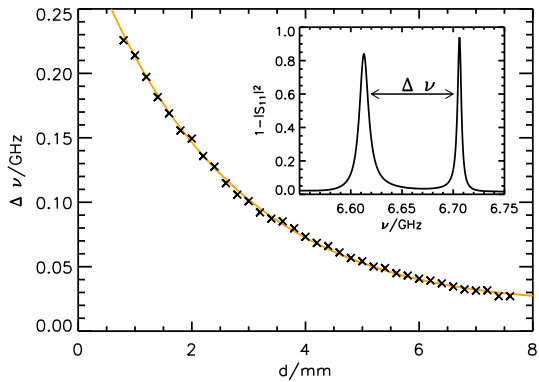}
\vspace{-0.4cm}

\caption{\label{fig:TwoDiscCoupling}
(color online) Splitting of the eigenfrequencies of two disks as a function of the distance caused by evanescent coupling. The experimental data (crosses) can be described by a Bessel function $K_0$ (see Eqs~(\ref{eq:kappad}) and (\ref{eq:deltanu})) where $\kappa_0\approx 1.82$\,GHz and $\gamma \approx 0.288$\,mm$^{-1}$ and a small offset $\Delta\nu_0 \approx$ 0.022\,GHz.
The inset shows the spectrum and splitting for $d$=3.3\,mm.
\vspace*{-0.5cm}}
\end{figure}

Now we investigate the coupling of two discs. Between the disks the TE$_1$ mode can be described approximately by only a $z$ component of the magnetic field $\vec{B}=(0,0,B_z)$ and perpendicular components of the electric field. Let us assume for the moment that both the bottom and top plates touch the disks from above and below. Then the $z$ dependence can be separated, yielding for the lowest TE-mode close to an individual disk \begin{equation}\label{eq:Bz}
B_z(x,y,z,k)=B_0 \times \left\{\begin{array}{cc}
\ \ \ \sin\left(\frac{\pi}{nh}z\right) J_0(k_\perp r) & r < r_d\\
\alpha \sin\left(\frac{\pi}{h}z\right) K_0(\gamma r) & r > r_d
\end{array}
\right.
\end{equation}
where $k_\perp=\sqrt{k^2-\left(\frac{\pi}{nh}\right)^2}$, $\gamma=\sqrt{\left(\frac{\pi}{h}\right)^2-k^2}$, and $r$ is the distance from the center of the disc. $J_0$ and $K_0$ are Bessel functions and $\alpha$ is a constant to be determined from the continuity equations at the surface. At the resonance frequency of the disk the wave number for $r>r_d$ is thus purely imaginary leading to an evanescent coupling between the discs.
Since in reality there is a gap between the top plate and the discs, this can be correct only approximately. But still we use Eq.~(\ref{eq:Bz}) with an effective $\gamma$ to be determined from the experiment.
A direct coupling to the TM$_0$ and TM$_1$ modes is strongly suppressed due to the continuity conditions of the fields. From Eq.~(\ref{eq:Bz}) we obtain for the coupling between two disks \begin{equation}
\label{eq:kappad}
\kappa(d)=\kappa_0 \left| K_0\left(\gamma\left[r_d+\frac{d}{2}\right]\right)\right|^2
\end{equation}
where $d$ is the distance between the disks ($d$=0 means the disks are touching) and $\kappa_0$ is a constant. The frequency splitting is given by
\begin{equation}
\label{eq:deltanu}
\Delta \nu(d) = \sqrt{4 \kappa(d)^2 + \left(\Delta\nu_0\right)^2},
\end{equation}
where $\Delta\nu_0$ is taking into account a small eigenfrequency difference of the disks and the differing coupling of the antenna to the discs, which actually dominates.
The resonance splitting is shown in Fig.~\ref{fig:TwoDiscCoupling} exhibiting an agreement with the predicted behavior. Thus each disk brings in just one bound state coupled evanescently to its neighbors, exactly the situation found in a tight-binding system.

\begin{figure}
\includegraphics[width=.8\columnwidth]{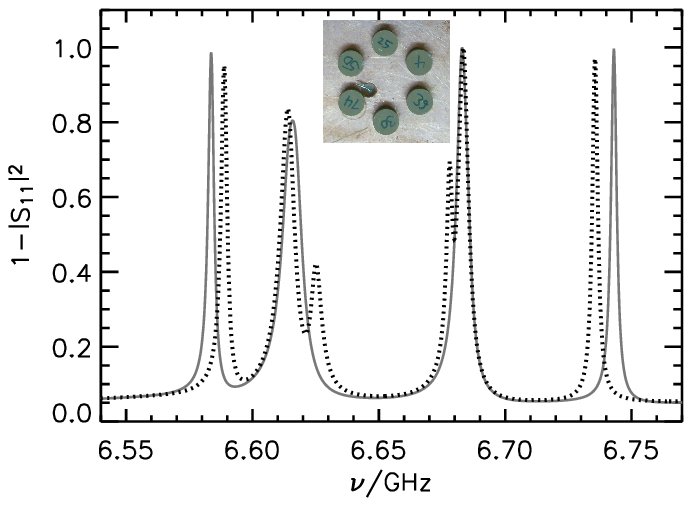}
\vspace{-0.cm}

\caption{\label{fig:Benzene}
(color online) Spectrum of a single benzene cell. The dotted line corresponds to a spectrum of a perturbed benzene cell, where the lifting of the degeneracy can be seen.
\vspace*{-0.5cm}}
\end{figure}

As a next test of the tight-binding approximation, we performed a measurement with three disks in a row as a function of the distance and found the coupling constant to be consistent with the two disk measurement. The next-nearest-neighbor coupling showed up to be about 6\% of the nearest neighbor coupling.
As graphene is made of coupled benzene cells we investigated the benzene cell (see inset of Fig.~\ref{fig:Benzene}) next. In Fig.~\ref{fig:Benzene} the spectrum (dark solid line) is shown. Benzene has dihedral symmetry D$_{6h}$ with two singlets and two doublets, where the singlets have the extremal energies. Correspondingly we observe four resonances. The degeneracy of the doublets can be lifted by a perturbation, e.\,g., by moving one disk (see dotted line). Again we performed a measurement varying the distance and found a good agreement with the coupling constant obtained by the two disk measurement.

Before we come to the final measurements we have to establish a connection between our measured quantity, the reflection $S_{11}$, and the DOS or more precisely the LDOS.
The starting point is the relation between the Green's function and the local density of states $\mathcal{L}(E)=-\frac{1}{\pi} \textrm{Im} G$.\cite{akk06}
Here we illustrate a short hand argument the connection between the reflection and the LDOS.
The total current through the antenna is given by $T=1-|S_{11}|^2= \oint d\textbf{s}\,\textbf{j}$, where $\textbf{j}=\textrm{Im}\Psi^*\nabla\Psi$ is the current density. The wave function of the field excited by a point-like antenna at $\mathbf{R}$ is proportional to the Green's function, $\psi \propto G(\mathbf{R},E)$,\cite{tud1}
\begin{eqnarray}
  \oint d\textbf{s}\,\textbf{j}&=&\frac{1}{2}\oint d\textbf{s}\,\left(\Psi\nabla\Psi^*-\Psi^*\nabla\Psi\right) \nonumber\\
  &=&-|B(\textbf{R};E)|^2\textrm{Im}\,G(\textbf{R};E).
\end{eqnarray}
where the proportionality coefficient $B(\mathbf{R};E)$ depends on the details of the coupling.
Thus finally we find
\begin{equation}\label{eq::mainrellong}
  1-|S_{11}(E)|^2=\pi |B(\textbf{R};E)|^2 \mathcal{L}(E),
\end{equation}
where $\mathcal{L}(E)$ is the LDOS of the unperturbed system.
To take into account the perturbation by the measurement antenna, in the correct derivation the Green's function $G$ has to be replaced by a properly renormalized Green's function $\xi_\beta$, where $\beta$ corresponds to the scattering length of the antenna.\cite{tud08,tud1} The coefficient $|B(\textbf{R};E)|^2$ is defined by the antenna properties which are slowly varying with energy and thus can be set to $|B(\textbf{R};E_D)|^2$ in the vicinity of the Dirac point $E_D=k_D^2$. With $\mathcal{L}(E)\sim|E-E_D|\sim|k-k_D|$ we expect
\begin{equation}\label{eq::mainrel}
  1-|S_{11}(E)|^2\sim |k-k_D|.
\end{equation}

\begin{figure}
\includegraphics[width=.9\columnwidth]{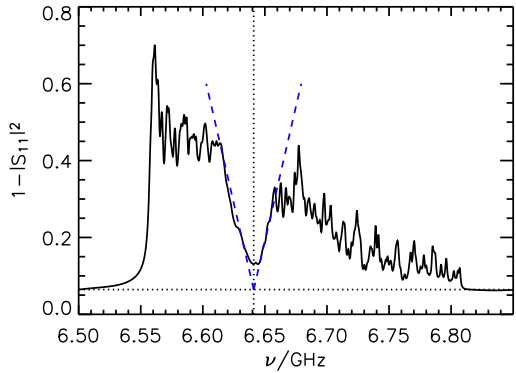}
\vspace{-0.5cm}

\caption{\label{fig:GrapheneInside}
(Color online) Spectrum of a graphene flake, where the spectrum was averaged over three different flake forms and different interior antenna positions. The dashed vertical line marks the resonance position of a single disk corresponding approximately to the Dirac point. The dotted horizontal line corresponds to the background. The dashed lines give a guide to the eye to see the linear increase of the DOS.
\vspace*{-0.5cm}}
\end{figure}

\begin{figure}
\includegraphics[width=.9\columnwidth]{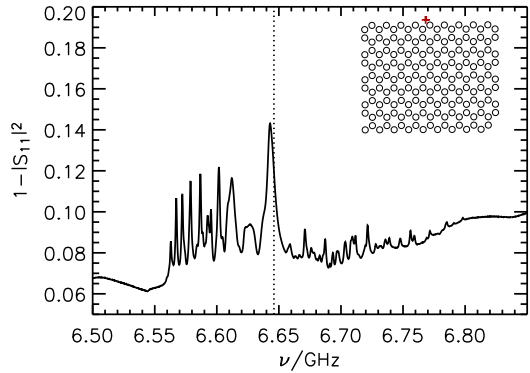}
\vspace{-0.5cm}

\caption{\label{fig:GrapheneEdge}
Spectrum of a graphene flake, where the antenna is placed at the zigzag boundary.
The inset shows the actual lattice and the cross marks the position of the measuring antenna.
The dashed vertical line is the resonance position of a single disc.
\vspace*{-0.4cm}}
\end{figure}

Now we can proceed to the discussion of our experimental findings for the graphene layers. We generated different graphene flakes and averaged over flakes and antenna positions. By this averaging procedure the LDOS is turned into the DOS. The result is shown in Fig.~\ref{fig:GrapheneInside}. At the frequency corresponding to the eigenfrequency a single disk (vertical dotted line) a spectral gap is observed which corresponds to the Dirac point in an infinite system. Additionally a symmetric linear increase in the DOS is found close to the Dirac point.
The overall decrease in $1-|S_{11}|^2$ with frequency has its origin in the fact that the measuring antenna is not placed at a disk center, but somewhere in between the discs, leading to a frequency dependent decrease in $B$ at the antenna due to Eq.~(\ref{eq:Bz}).

Now let us have a look what happens if we move the antenna to the outside, e.\,g.\ to the zigzag edge. This measurement is presented in Fig.~\ref{fig:GrapheneEdge}. A state close to the Dirac point is observed, which cannot be seen in Fig.~\ref{fig:GrapheneInside}, even though it is the same system.
As the measured quantity is related to the LDOS and not the DOS it is possible to test the states locally and not only globally. We also moved the antenna to the armchair edge and did not observe a state close to the Dirac point. This is in accordance with the findings for a zigzag-edged infinitely long graphene ribbons where the effect of edge states was discussed,\cite{mun06,cas09} and where the localization of the edge states was estimated to about one to two atomic layers. If we move the antenna to the lower left corner (see inset Fig.~\ref{fig:GrapheneEdge}) where there is an isolated disk not contained in a full benzene ring, again a single state is observed at the Dirac point. By removing the disk at the corner the state disappears.

\begin{figure}
\includegraphics[width=.9\columnwidth]{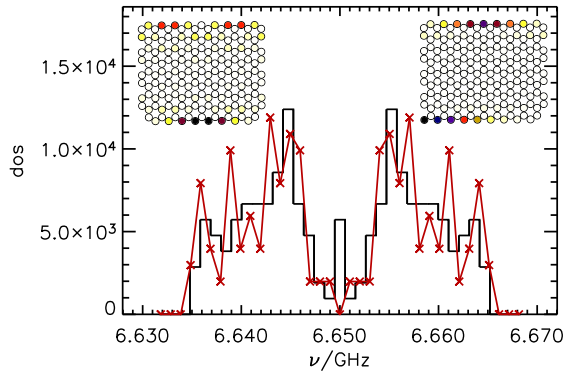}
\vspace{-0.5cm}

\caption{\label{fig:NumDOS_Eigenfunction}
(color online) DOS from a numerical calculation of the systems shown in the top right corner (histogram). Additionally the right inset shows an grey scale indication of an edge state found close to the Dirac frequency.
In the left inset an typical eigenfunction of an edge state is presented, where the disk in the lower left corner was removed.
The crosses correspond to a system with periodic boundary conditions taking 178 disks into account.
\vspace*{-0.5cm}}
\end{figure}

To interpret our experimental findings we additionally performed numerical calculations by
determining eigenvalues and eigenvectors of the Hamiltonian
\begin{equation}
H=E_D\cdot 1 +\kappa C,
\end{equation}
where $E_D$ is the gap energy, $\kappa$ the coupling constant, and $C$ the connectivity matrix with elements $C_{nm}=$\,1 or 0 for nearest and non-nearest neighbors, respectively. The resulting DOS is shown in Fig.~\ref{fig:NumDOS_Eigenfunction}. The histogram corresponds to the experimental findings presented in Fig.~\ref{fig:GrapheneEdge}, where the edge states are strongly observed but internal states are suppressed. One observes a peak of the DOS at the Dirac point. Analyzing the wave functions at this energy we found only edge states similar to the one presented in the left inset of Fig.~\ref{fig:NumDOS_Eigenfunction} as well as a single corner state observed if an individual disk not part of a full benzene cell is added (see right inset).
By moving the measurement antenna toward the center of the system, corner or edge states cannot be seen as their probability inside is very small and the measured quantity is related to the LDOS.
In addition we performed the calculation with periodic boundary conditions. Now the edge states have disappeared, and the triangular gap at the Dirac point is clearly seen (crosses in Fig.~\ref{fig:NumDOS_Eigenfunction}). This is similar to the situation of an internal antenna, i.\,e., it corresponds to Fig.~\ref{fig:GrapheneInside}, where the Dirac point is found as well. In the numerics with periodic boundary conditions edge states can not exist, whereas in the experiment the existing edge states are not observable, due to their small amplitude inside the flake.
Finally by removing the corner disk leaving the lattice only with complete benzene cells the corner state did vanished, again in accordance with the experimental findings.

In this paper we presented the first microwave realization of graphene in the tight-binding approximation. Using high index of refraction disks inside two metallic plates, where the disks resonances are only coupled evanescently, a tight-binding realization became possible. The Dirac point and a linear increase in the DOS close to it have been found as well as edge and corner states. It was essential for this approach to conceive a system where each disk brings in just one state coupling evanescently with its neighbors. This is a qualitatively new feature for the microwave experiments as a whole, enabling a new class of experiments in the context of molecular orbital theory, going by far beyond the specific example of graphene studied in the present work. The results for benzene shown in Fig.~\ref{fig:Benzene} illustrate for a textbook example the feasibility of the method. Due to the flexibility of the set-up and the closeness of the system additionally graphene like systems with disorder, defects, graphene quantum dots \cite{lib09} or even Dirac oscillators \cite{sad10} can be investigated. By introducing different types of disks with different resonance frequencies it is also possible to investigate systems with different atom per unit celllike boron-nitride, corresponding to a Dirac equation for particles with masses.

This work was supported by the Deutsche Forschungsgemeinschaft via an individual grant and the Forschergruppe 760: Scattering systems with complex dynamics.
U.~K.\ and S.~B.\ thank the LPMC at Nice for the hospitality during different long term visits
and the University of Nice and the F\'{e}d\'{e}ration D\"{o}blin for financial supports.
We thank M. Miski-Oglu, T.~Seligman, and E.~Sadurn\'{\i} for helpful discussions.

\vspace*{-0.5cm}

\begin{thebibliography}{22}%
\makeatletter
\providecommand \@ifxundefined [1]{%
 \@ifx{#1\undefined}
}%
\providecommand \@ifnum [1]{%
 \ifnum #1\expandafter \@firstoftwo
 \else \expandafter \@secondoftwo
 \fi
}%
\providecommand \@ifx [1]{%
 \ifx #1\expandafter \@firstoftwo
 \else \expandafter \@secondoftwo
 \fi
}%
\providecommand \natexlab [1]{#1}%
\providecommand \enquote  [1]{``#1''}%
\providecommand \bibnamefont  [1]{#1}%
\providecommand \bibfnamefont [1]{#1}%
\providecommand \citenamefont [1]{#1}%
\providecommand \href@noop [0]{\@secondoftwo}%
\providecommand \href [0]{\begingroup \@sanitize@url \@href}%
\providecommand \@href[1]{\@@startlink{#1}\@@href}%
\providecommand \@@href[1]{\endgroup#1\@@endlink}%
\providecommand \@sanitize@url [0]{\catcode `\\12\catcode `\$12\catcode
  `\&12\catcode `\#12\catcode `\^12\catcode `\_12\catcode `\%12\relax}%
\providecommand \@@startlink[1]{}%
\providecommand \@@endlink[0]{}%
\providecommand \url  [0]{\begingroup\@sanitize@url \@url }%
\providecommand \@url [1]{\endgroup\@href {#1}{\urlprefix }}%
\providecommand \urlprefix  [0]{URL }%
\providecommand \Eprint [0]{\href }%
\@ifxundefined \urlstyle {%
  \providecommand \doi  [0]{\begingroup \@sanitize@url \@doi}%
  \providecommand \@doi [1]{\endgroup \@@startlink {\doibase
  #1}doi:\discretionary {}{}{}#1\@@endlink }%
}{%
  \providecommand \doi  [0]{doi:\discretionary{}{}{}\begingroup
  \urlstyle{rm}\Url }%
}%
\providecommand \doibase [0]{http://dx.doi.org/}%
\providecommand \Doi [0]{\begingroup \@sanitize@url \@Doi }%
\providecommand \@Doi  [1]{\endgroup\@@startlink{\doibase#1}\@@Doi}%
\providecommand \@@Doi [1]{#1\@@endlink}%
\providecommand \selectlanguage [0]{\@gobble}%
\providecommand \bibinfo  [0]{\@secondoftwo}%
\providecommand \bibfield  [0]{\@secondoftwo}%
\providecommand \translation [1]{[#1]}%
\providecommand \BibitemOpen [0]{}%
\providecommand \bibitemStop [0]{}%
\providecommand \bibitemNoStop [0]{.\EOS\space}%
\providecommand \EOS [0]{\spacefactor3000\relax}%
\providecommand \BibitemShut  [1]{\csname bibitem#1\endcsname}%
\bibitem [{\citenamefont {Wallace}(1947)}]{wal47}%
  \BibitemOpen
  \bibfield  {author} {\bibinfo {author} {\bibfnamefont {P.~R.}\ \bibnamefont
  {Wallace}},\ }\href@noop {} {\bibfield  {journal} {\bibinfo  {journal} {Phys.
  Rev.},\ }\textbf {\bibinfo {volume} {71}},\ \bibinfo {pages} {622} (\bibinfo
  {year} {1947})}\BibitemShut {NoStop}%
\bibitem [{\citenamefont {Semenoff}(1984)}]{sem84}%
  \BibitemOpen
  \bibfield  {author} {\bibinfo {author} {\bibfnamefont {G.~W.}\ \bibnamefont
  {Semenoff}},\ }\href@noop {} {\bibfield  {journal} {\bibinfo  {journal}
  {Phys. Rev. Lett.},\ }\textbf {\bibinfo {volume} {53}},\ \bibinfo {pages}
  {2449} (\bibinfo {year} {1984})}\BibitemShut {NoStop}%
\bibitem [{\citenamefont {Novoselov}\ \emph {et~al.}(2005)\citenamefont
  {Novoselov}, \citenamefont {Geim}, \citenamefont {Morozov}, \citenamefont
  {Jiang}, \citenamefont {Katsnelson}, \citenamefont {Grigorieva},
  \citenamefont {Dubonos},\ and\ \citenamefont {Firsov}}]{nov05}%
  \BibitemOpen
  \bibfield  {author} {\bibinfo {author} {\bibfnamefont {K.~S.}\ \bibnamefont
  {Novoselov}}, \bibinfo {author} {\bibfnamefont {A.~K.}\ \bibnamefont {Geim}},
  \bibinfo {author} {\bibfnamefont {S.~V.}\ \bibnamefont {Morozov}}, \bibinfo
  {author} {\bibfnamefont {D.}~\bibnamefont {Jiang}}, \bibinfo {author}
  {\bibfnamefont {M.~I.}\ \bibnamefont {Katsnelson}}, \bibinfo {author}
  {\bibfnamefont {I.~V.}\ \bibnamefont {Grigorieva}}, \bibinfo {author}
  {\bibfnamefont {S.~V.}\ \bibnamefont {Dubonos}}, \ and\ \bibinfo {author}
  {\bibfnamefont {A.~A.}\ \bibnamefont {Firsov}},\ }\Doi
  {doi:10.1038/nature04233} {\bibfield  {journal} {\bibinfo  {journal}
  {Nature},\ }\textbf {\bibinfo {volume} {438}},\ \bibinfo {pages} {197}
  (\bibinfo {year} {2005})}\BibitemShut {NoStop}%
\bibitem [{\citenamefont {{Castro Neto}}\ \emph {et~al.}(2009)\citenamefont
  {{Castro Neto}}, \citenamefont {Guinea}, \citenamefont {Peres}, \citenamefont
  {Novoselov},\ and\ \citenamefont {Geim}}]{cas09}%
  \BibitemOpen
  \bibfield  {author} {\bibinfo {author} {\bibfnamefont {A.~H.}\ \bibnamefont
  {{Castro Neto}}}, \bibinfo {author} {\bibfnamefont {F.}~\bibnamefont
  {Guinea}}, \bibinfo {author} {\bibfnamefont {N.~M.~R.}\ \bibnamefont
  {Peres}}, \bibinfo {author} {\bibfnamefont {K.~S.}\ \bibnamefont
  {Novoselov}}, \ and\ \bibinfo {author} {\bibfnamefont {A.~K.}\ \bibnamefont
  {Geim}},\ }\href@noop {} {\bibfield  {journal} {\bibinfo  {journal} {Rev.
  Mod. Phys.},\ }\textbf {\bibinfo {volume} {81}},\ \bibinfo {pages} {109}
  (\bibinfo {year} {2009})}\BibitemShut {NoStop}%
\bibitem [{\citenamefont {Novoselov}\ \emph {et~al.}(2004)\citenamefont
  {Novoselov}, \citenamefont {Geim}, \citenamefont {Morozov}, \citenamefont
  {Jiang}, \citenamefont {Zhang}, \citenamefont {Dubonos}, \citenamefont
  {Grigorieva},\ and\ \citenamefont {Firsov}}]{nov04}%
  \BibitemOpen
  \bibfield  {author} {\bibinfo {author} {\bibfnamefont {K.~S.}\ \bibnamefont
  {Novoselov}}, \bibinfo {author} {\bibfnamefont {A.~K.}\ \bibnamefont {Geim}},
  \bibinfo {author} {\bibfnamefont {S.~V.}\ \bibnamefont {Morozov}}, \bibinfo
  {author} {\bibfnamefont {D.}~\bibnamefont {Jiang}}, \bibinfo {author}
  {\bibfnamefont {Y.}~\bibnamefont {Zhang}}, \bibinfo {author} {\bibfnamefont
  {S.~V.}\ \bibnamefont {Dubonos}}, \bibinfo {author} {\bibfnamefont {I.~V.}\
  \bibnamefont {Grigorieva}}, \ and\ \bibinfo {author} {\bibfnamefont {A.~A.}\
  \bibnamefont {Firsov}},\ }\href@noop {} {\bibfield  {journal} {\bibinfo
  {journal} {Science},\ }\textbf {\bibinfo {volume} {306}},\ \bibinfo {pages}
  {666} (\bibinfo {year} {2004})}\BibitemShut {NoStop}%
\bibitem [{\citenamefont {Geim}\ and\ \citenamefont {Novoselov}(2007)}]{gei07}%
  \BibitemOpen
  \bibfield  {author} {\bibinfo {author} {\bibfnamefont {A.~K.}\ \bibnamefont
  {Geim}}\ and\ \bibinfo {author} {\bibfnamefont {K.~S.}\ \bibnamefont
  {Novoselov}},\ }\Doi {doi:10.1038/nature04233} {\bibfield  {journal}
  {\bibinfo  {journal} {Nature Materials},\ }\textbf {\bibinfo {volume} {6}},\
  \bibinfo {pages} {183} (\bibinfo {year} {2007})}\BibitemShut {NoStop}%
\bibitem [{\citenamefont {Soukoulis}(1993)}]{sou93}%
  \BibitemOpen
  \bibinfo {editor} {\bibfnamefont {C.~M.}\ \bibnamefont {Soukoulis}},\ ed.,\
  \href@noop {} {\emph {\bibinfo {title} {Photonic Band Gaps and
  Localization}}},\ NATO ASI Series B: Physics, Vol. 308\ (\bibinfo
  {publisher} {Plenum Press},\ \bibinfo {address} {New York},\ \bibinfo {year}
  {1993})\BibitemShut {NoStop}%
\bibitem [{\citenamefont {Lourtioz}\ \emph {et~al.}(2005)\citenamefont
  {Lourtioz}, \citenamefont {Benisty}, \citenamefont {Berger}, \citenamefont
  {Gerard}, \citenamefont {Maystre},\ and\ \citenamefont {Tchelnokov}}]{lou05}%
  \BibitemOpen
  \bibfield  {author} {\bibinfo {author} {\bibfnamefont {J.~M.}\ \bibnamefont
  {Lourtioz}}, \bibinfo {author} {\bibfnamefont {H.}~\bibnamefont {Benisty}},
  \bibinfo {author} {\bibfnamefont {V.}~\bibnamefont {Berger}}, \bibinfo
  {author} {\bibfnamefont {J.~M.}\ \bibnamefont {Gerard}}, \bibinfo {author}
  {\bibfnamefont {D.}~\bibnamefont {Maystre}}, \ and\ \bibinfo {author}
  {\bibfnamefont {A.}~\bibnamefont {Tchelnokov}},\ }\href@noop {} {\emph
  {\bibinfo {title} {Photonic Crystals - Towards Nanoscale Photonic Devices}}}\
  (\bibinfo  {publisher} {Springer},\ \bibinfo {address} {Berlin},\ \bibinfo
  {year} {2005})\BibitemShut {NoStop}%
\bibitem [{\citenamefont {Sepkhanov}\ \emph {et~al.}(2008)\citenamefont
  {Sepkhanov}, \citenamefont {Nilsson},\ and\ \citenamefont
  {Beenakker}}]{sep08}%
  \BibitemOpen
  \bibfield  {author} {\bibinfo {author} {\bibfnamefont {R.~A.}\ \bibnamefont
  {Sepkhanov}}, \bibinfo {author} {\bibfnamefont {J.}~\bibnamefont {Nilsson}},
  \ and\ \bibinfo {author} {\bibfnamefont {C.~W.~J.}\ \bibnamefont
  {Beenakker}},\ }\href@noop {} {\bibfield  {journal} {\bibinfo  {journal}
  {Phys. Rev. B},\ }\textbf {\bibinfo {volume} {78}},\ \bibinfo {pages}
  {045122} (\bibinfo {year} {2008})}\BibitemShut {NoStop}%
\bibitem [{\citenamefont {Zandbergen}\ and\ \citenamefont
  {de~Dood}(2010)}]{zan10}%
  \BibitemOpen
  \bibfield  {author} {\bibinfo {author} {\bibfnamefont {S.~R.}\ \bibnamefont
  {Zandbergen}}\ and\ \bibinfo {author} {\bibfnamefont {M.~J.~A.}\ \bibnamefont
  {de~Dood}},\ }\href@noop {} {\bibfield  {journal} {\bibinfo  {journal} {Phys.
  Rev. Lett.},\ }\textbf {\bibinfo {volume} {104}},\ \bibinfo {pages} {043903}
  (\bibinfo {year} {2010})}\BibitemShut {NoStop}%
\bibitem [{\citenamefont {Bittner}\ \emph {et~al.}(2010)\citenamefont
  {Bittner}, \citenamefont {Dietz}, \citenamefont {Miski-Oglu}, \citenamefont
  {Iriarte}, \citenamefont {Richter},\ and\ \citenamefont
  {Sch{\"a}fer}}]{bit10}%
  \BibitemOpen
  \bibfield  {author} {\bibinfo {author} {\bibfnamefont {S.}~\bibnamefont
  {Bittner}}, \bibinfo {author} {\bibfnamefont {B.}~\bibnamefont {Dietz}},
  \bibinfo {author} {\bibfnamefont {M.}~\bibnamefont {Miski-Oglu}}, \bibinfo
  {author} {\bibfnamefont {P.~O.}\ \bibnamefont {Iriarte}}, \bibinfo {author}
  {\bibfnamefont {A.}~\bibnamefont {Richter}}, \ and\ \bibinfo {author}
  {\bibfnamefont {F.}~\bibnamefont {Sch{\"a}fer}},\ }\href@noop {} {\bibfield
  {journal} {\bibinfo  {journal} {Phys. Rev. B},\ }\textbf {\bibinfo {volume}
  {82}},\ \bibinfo {pages} {014301} (\bibinfo {year} {2010})}\BibitemShut
  {NoStop}%
\bibitem [{\citenamefont {Bena}\ and\ \citenamefont
  {Montambaux}(2009)}]{ben09b}%
  \BibitemOpen
  \bibfield  {author} {\bibinfo {author} {\bibfnamefont {C.}~\bibnamefont
  {Bena}}\ and\ \bibinfo {author} {\bibfnamefont {G.}~\bibnamefont
  {Montambaux}},\ }\href@noop {} {\bibfield  {journal} {\bibinfo  {journal}
  {New J. of Physics},\ }\textbf {\bibinfo {volume} {11}},\ \bibinfo {pages}
  {095003} (\bibinfo {year} {2009})}\BibitemShut {NoStop}%
\bibitem [{\citenamefont {Kuhl}\ and\ \citenamefont
  {St{\"o}ckmann}(1998)}]{kuh98b}%
  \BibitemOpen
  \bibfield  {author} {\bibinfo {author} {\bibfnamefont {U.}~\bibnamefont
  {Kuhl}}\ and\ \bibinfo {author} {\bibfnamefont {H.-J.}\ \bibnamefont
  {St{\"o}ckmann}},\ }\href@noop {} {\bibfield  {journal} {\bibinfo  {journal}
  {Phys. Rev. Lett.},\ }\textbf {\bibinfo {volume} {80}},\ \bibinfo {pages}
  {3232} (\bibinfo {year} {1998})}\BibitemShut {NoStop}%
\bibitem [{\citenamefont {Chabanov}\ \emph {et~al.}(2000)\citenamefont
  {Chabanov}, \citenamefont {Stoytchev},\ and\ \citenamefont {Genack}}]{cha00}%
  \BibitemOpen
  \bibfield  {author} {\bibinfo {author} {\bibfnamefont {A.~A.}\ \bibnamefont
  {Chabanov}}, \bibinfo {author} {\bibfnamefont {M.}~\bibnamefont {Stoytchev}},
  \ and\ \bibinfo {author} {\bibfnamefont {A.~Z.}\ \bibnamefont {Genack}},\
  }\href@noop {} {\bibfield  {journal} {\bibinfo  {journal} {Nature},\ }\textbf
  {\bibinfo {volume} {404}},\ \bibinfo {pages} {850} (\bibinfo {year}
  {2000})}\BibitemShut {NoStop}%
\bibitem [{\citenamefont {Laurent}\ \emph {et~al.}(2007)\citenamefont
  {Laurent}, \citenamefont {Legrand}, \citenamefont {Sebbah}, \citenamefont
  {Vanneste},\ and\ \citenamefont {Mortessagne}}]{lau07}%
  \BibitemOpen
  \bibfield  {author} {\bibinfo {author} {\bibfnamefont {D.}~\bibnamefont
  {Laurent}}, \bibinfo {author} {\bibfnamefont {O.}~\bibnamefont {Legrand}},
  \bibinfo {author} {\bibfnamefont {P.}~\bibnamefont {Sebbah}}, \bibinfo
  {author} {\bibfnamefont {C.}~\bibnamefont {Vanneste}}, \ and\ \bibinfo
  {author} {\bibfnamefont {F.}~\bibnamefont {Mortessagne}},\ }\href@noop {}
  {\bibfield  {journal} {\bibinfo  {journal} {Phys. Rev. Lett.},\ }\textbf
  {\bibinfo {volume} {99}},\ \bibinfo {pages} {253902} (\bibinfo {year}
  {2007})}\BibitemShut {NoStop}%
\bibitem [{\citenamefont {Kuhl}\ \emph {et~al.}(2008)\citenamefont {Kuhl},
  \citenamefont {Izrailev},\ and\ \citenamefont {Krokhin}}]{kuh08a}%
  \BibitemOpen
  \bibfield  {author} {\bibinfo {author} {\bibfnamefont {U.}~\bibnamefont
  {Kuhl}}, \bibinfo {author} {\bibfnamefont {F.~M.}\ \bibnamefont {Izrailev}},
  \ and\ \bibinfo {author} {\bibfnamefont {A.~A.}\ \bibnamefont {Krokhin}},\
  }\href@noop {} {\bibfield  {journal} {\bibinfo  {journal} {Phys. Rev.
  Lett.},\ }\textbf {\bibinfo {volume} {100}},\ \bibinfo {pages} {126402}
  (\bibinfo {year} {2008})}\BibitemShut {NoStop}%
\bibitem [{\citenamefont {Akkermans}\ and\ \citenamefont
  {Montambaux}(2006)}]{akk06}%
  \BibitemOpen
  \bibinfo {editor} {\bibfnamefont {E.}~\bibnamefont {Akkermans}}\ and\
  \bibinfo {editor} {\bibfnamefont {G.}~\bibnamefont {Montambaux}},\ eds.,\
  \href@noop {} {\emph {\bibinfo {title} {Mesoscopic Physics of electrons and
  photons}}}\ (\bibinfo  {publisher} {Cambridge University Press},\ \bibinfo
  {address} {Cambridge},\ \bibinfo {year} {2006})\BibitemShut {NoStop}%
\bibitem [{\citenamefont {Tudorovskiy}\ \emph {et~al.}(2009)\citenamefont
  {Tudorovskiy}, \citenamefont {Kuhl},\ and\ \citenamefont
  {St{\"o}ckmann}}]{tud1}%
  \BibitemOpen
  \bibfield  {author} {\bibinfo {author} {\bibfnamefont {T.}~\bibnamefont
  {Tudorovskiy}}, \bibinfo {author} {\bibfnamefont {U.}~\bibnamefont {Kuhl}}, \
  and\ \bibinfo {author} {\bibfnamefont {H.-J.}\ \bibnamefont
  {St{\"o}ckmann}},\ }\href@noop {} {\enquote {\bibinfo {title} {Singular
  statistics revised},}\ }\bibinfo {howpublished} {Preprint} (\bibinfo {year}
  {2009}),\ \bibinfo {note} {arXiv:0910.3079}\BibitemShut {NoStop}%
\bibitem [{\citenamefont {Tudorovskiy}\ \emph {et~al.}(2008)\citenamefont
  {Tudorovskiy}, \citenamefont {H{\"o}hmann}, \citenamefont {Kuhl},\ and\
  \citenamefont {St{\"o}ckmann}}]{tud08}%
  \BibitemOpen
  \bibfield  {author} {\bibinfo {author} {\bibfnamefont {T.}~\bibnamefont
  {Tudorovskiy}}, \bibinfo {author} {\bibfnamefont {R.}~\bibnamefont
  {H{\"o}hmann}}, \bibinfo {author} {\bibfnamefont {U.}~\bibnamefont {Kuhl}}, \
  and\ \bibinfo {author} {\bibfnamefont {H.-J.}\ \bibnamefont
  {St{\"o}ckmann}},\ }\href@noop {} {\bibfield  {journal} {\bibinfo  {journal}
  {J. Phys. A},\ }\textbf {\bibinfo {volume} {41}},\ \bibinfo {pages} {275101}
  (\bibinfo {year} {2008})}\BibitemShut {NoStop}%
\bibitem [{\citenamefont {Mu{\~n}oz-Rojas}\ \emph {et~al.}(2006)\citenamefont
  {Mu{\~n}oz-Rojas}, \citenamefont {Jacob}, \citenamefont
  {Fern{\'a}ndez-Rossier},\ and\ \citenamefont {Palacios}}]{mun06}%
  \BibitemOpen
  \bibfield  {author} {\bibinfo {author} {\bibfnamefont {F.}~\bibnamefont
  {Mu{\~n}oz-Rojas}}, \bibinfo {author} {\bibfnamefont {D.}~\bibnamefont
  {Jacob}}, \bibinfo {author} {\bibfnamefont {J.}~\bibnamefont
  {Fern{\'a}ndez-Rossier}}, \ and\ \bibinfo {author} {\bibfnamefont {J.~J.}\
  \bibnamefont {Palacios}},\ }\href@noop {} {\bibfield  {journal} {\bibinfo
  {journal} {Phys. Rev. B},\ }\textbf {\bibinfo {volume} {74}},\ \bibinfo
  {pages} {195417} (\bibinfo {year} {2006})}\BibitemShut {NoStop}%
\bibitem [{\citenamefont {Libisch}\ \emph {et~al.}(2009)\citenamefont
  {Libisch}, \citenamefont {Stampfer},\ and\ \citenamefont
  {Burgd{\"o}rfer}}]{lib09}%
  \BibitemOpen
  \bibfield  {author} {\bibinfo {author} {\bibfnamefont {F.}~\bibnamefont
  {Libisch}}, \bibinfo {author} {\bibfnamefont {C.}~\bibnamefont {Stampfer}}, \
  and\ \bibinfo {author} {\bibfnamefont {J.}~\bibnamefont {Burgd{\"o}rfer}},\
  }\href@noop {} {\bibfield  {journal} {\bibinfo  {journal} {Phys. Rev. B},\
  }\textbf {\bibinfo {volume} {79}},\ \bibinfo {pages} {115423} (\bibinfo
  {year} {2009})}\BibitemShut {NoStop}%
\bibitem [{\citenamefont {Sadurn{\'\i}}\ \emph {et~al.}(2010)\citenamefont
  {Sadurn{\'\i}}, \citenamefont {Seligman},\ and\ \citenamefont
  {Mortessagne}}]{sad10}%
  \BibitemOpen
  \bibfield  {author} {\bibinfo {author} {\bibfnamefont {E.}~\bibnamefont
  {Sadurn{\'\i}}}, \bibinfo {author} {\bibfnamefont {T.}~\bibnamefont
  {Seligman}}, \ and\ \bibinfo {author} {\bibfnamefont {F.}~\bibnamefont
  {Mortessagne}},\ }\href@noop {} {\bibfield  {journal} {\bibinfo  {journal}
  {New J. of Physics},\ }\textbf {\bibinfo {volume} {12}},\ \bibinfo {pages}
  {053014} (\bibinfo {year} {2010})}\BibitemShut {NoStop}%
\end{thebibliography}
\end{document}